\documentclass[conference]{IEEEtran}
\IEEEoverridecommandlockouts
\usepackage{cite}
\usepackage{amsmath,amssymb,amsfonts}
\usepackage{algorithmic}
\usepackage{graphicx}
\usepackage{textcomp}
\usepackage{xcolor}
\usepackage{algorithmic}
\usepackage{graphicx,wrapfig,lipsum}
\usepackage{color}
\usepackage{comment}

\usepackage[ruled,vlined]{algorithm2e}
\def\BibTeX{{\rm B\kern-.05em{\sc i\kern-.025em b}\kern-.08em
    T\kern-.1667em\lower.7ex\hbox{E}\kern-.125emX}}
\begin{document}

\title{EMSCA \& FI Self-Awareness and Resilience with Single On-chip Loop \& FCN Classifier}

\title{A $334uW$ $0.158mm^2$ Saber Learning with Rounding based Post-Quantum Crypto Accelerator}

\title{A $333.9uW$ $0.158mm^2$ Saber Learning with Rounding based Post-Quantum Crypto Accelerator}

\author{\IEEEauthorblockA{Archisman Ghosh\IEEEauthorrefmark{1},
J.M.B. Mera\IEEEauthorrefmark{3},
Angshuman Karmakar\IEEEauthorrefmark{3},
Debayan Das\IEEEauthorrefmark{1},
Santosh Ghosh\IEEEauthorrefmark{2},
Ingrid Verbauwhede\IEEEauthorrefmark{3}
and~Shreyas~Sen\IEEEauthorrefmark{1}} 
\IEEEauthorblockA{\IEEEauthorrefmark{1}School of Electrical and Computer Engineering,
Purdue University, West Lafayette, IN, USA} 
\IEEEauthorblockA{\IEEEauthorrefmark{3}COSIC, KU Leuven, Belgium}
\IEEEauthorblockA{\IEEEauthorrefmark{2}Intel Labs, Intel Corporation, Hillsboro, OR. USA}
}
\maketitle
\vspace{-5mm}
\begin{abstract}
National Institute of Standard \& Technology (NIST) is currently running a multi\-year\-long standardization procedure to select quantum\-safe or post\-quantum cryptographic schemes to be used in the future. Saber is the only LWR based algorithm to be in the final of Round 3. This work presents a Saber ASIC which provides 1.37X power-efficient, 1.75$\times$ lower area, and 4$\times$ less memory implementation w.r.t. other SoA PQC ASIC. The energy\-hungry multiplier block is 1.5$\times$ energy\-efficient than SoA. 
\end{abstract}


\vspace{-1mm}
\section{Introduction}
\vspace{-1mm}
The arrival of large-scale quantum computers will break the security assurances of our current public-key cryptography. National Institute of Standard \& Technology (NIST) is currently running a multi-year-long standardization procedure to select quantum-safe or post-quantum cryptographic schemes to be used in the future. Energy efficiency is an important criterion in the selection process. This paper presents the first Silicon verified ASIC implementation for Saber (LWR algorithm as proposed in \cite{cryptoeprint:2018:230}, \cite{banerjee2012pseudorandom}), a NIST PQC Round 3 finalist candidate in the key-encapsulation mechanism (KEM) category. Fig. 1 briefly describes the learning with rounding (LWR) problem, which is hard to solve even in the presence of large quantum computers due to the noise generated from rounding. IC features are tabulated in Fig. 1. which also shows a simplified version of the Saber KEM scheme to establish a secret key between two communicating parties Alice and Bob. Due to learning with rounding, secret s is hard to guess based on publicly available data as shown in Fig.1.

\begin{figure}[!t]
\centering
\includegraphics[width=1\columnwidth]{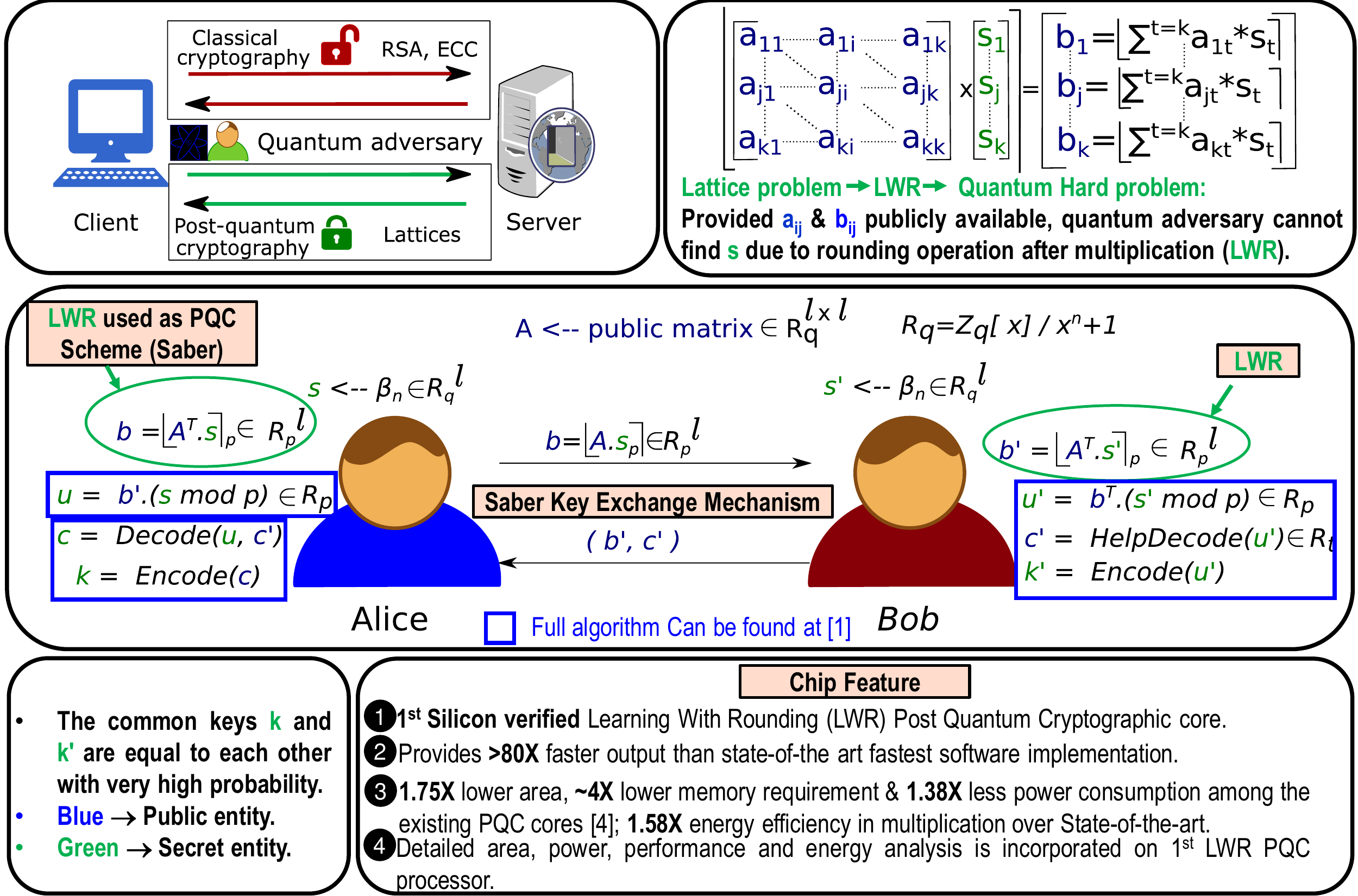}
\caption{Quantum resistant cryptographic core for IoT; Learning with rounding approach in lattice-based crypto; Key Exchange Mechanism (KEM) for a LWR based crypto algorithm (Saber); Chip features.}
\label{intro}
\vspace{-6mm}
\end{figure}

\vspace{-1mm}
\section{Architecture}
\vspace{-2mm}
\begin{figure*}
\centering
\includegraphics[width=0.9\textwidth]{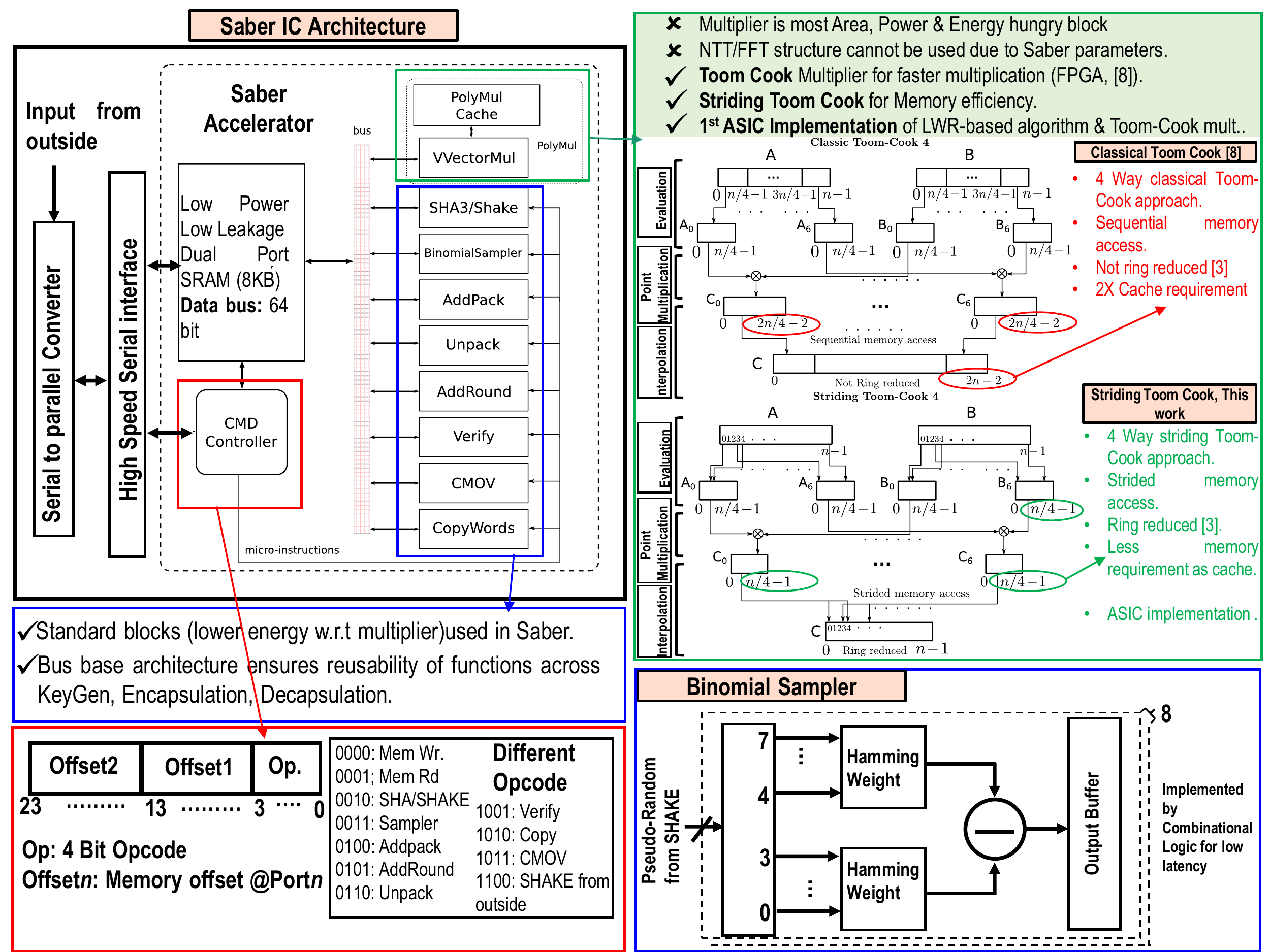}
\caption{Saber (NIST LWR based PQC Key Encapsulation Mechanism finalist) co-processor; New Toom-Cook multiplication architecture; OpCode format; Vector Sampler.}
\label{chip_micro}
\end{figure*}
\begin{figure*}[!t]
\centering
\includegraphics[width=2\columnwidth]{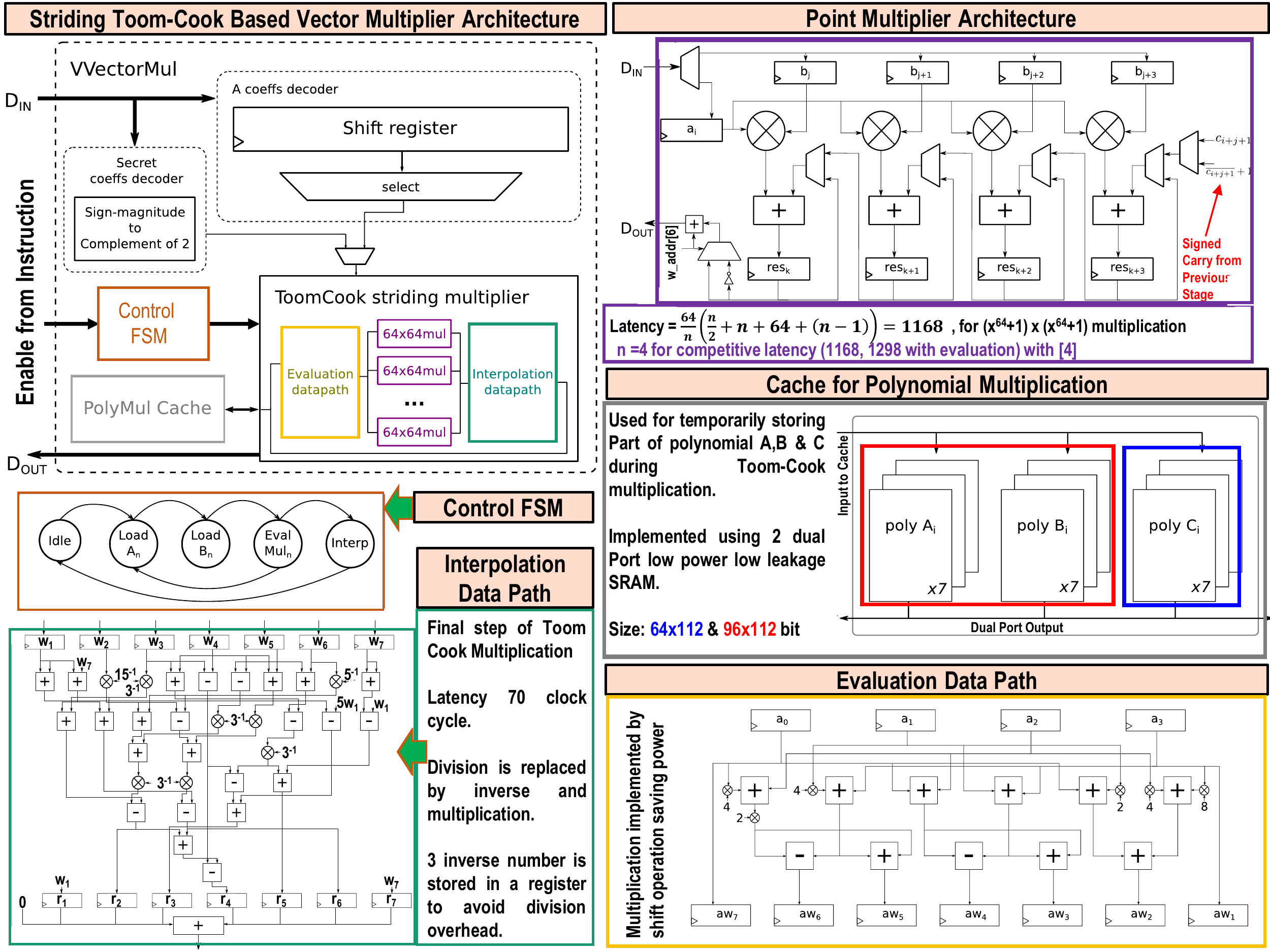}
\caption{Vector multiplier architecture; Control FSM for multiplier; Interpolation data path; Point multiplication; Cache access; Evaluation data path.}
\label{attack_setup}
\end{figure*}
Fig. 2 represents the architecture with self-explanatory named blocks to compute KeyGeneration, Encapsulation and Decapsulation of Saber. The blocks in the full chip architecture are equivalent to other lattice-based schemes, each of the operations involved in lattice-based cryptography are supported. The key innovation lies in the 1) multiplication and its implementation 2) the memory management and 3) interface of the co-processor. 1) Computationally, the most expensive and complex part of Saber is the multiplication of polynomials of length 256. Fastest method for such type of multiplication is the number theoretic transform (NTT), a special form of fast Fourier transforms for finite fields (used in other PQC algorithm [4]). However, Saber’s parameters do not directly satisfy the above condition for applying NTT. Saber with Toom-Cook and Karatsuba based multiplications have shown that implementations using these algorithms can be as fast as NTT multiplications. A brief Toom-Cook multiplication is shown in Fig. 2 along with the complete architecture. Since Saber needs the final result C(x) to be reduced by $x^{256}+1$, striding Toom-Cook multiplication is adapted for memory efficiency (In this case, instead of splitting the polynomial A(x) (B(x) similarly) as $A_0(x)= a_0+a_1x+. . . + a_{63}x^{63}$, $A_1(x)= a_{64}+a_{65}x+. . . + a_{127}x^{63}$,. . . ,  $A_3(x)= a_{192}+a_{193}x+. . . + a_{255}x^{63}$ during the evaluation stage of classical Toom-Cook multiplication, the polynomial A(x) is split with striding of 4 as $A_0(x)= a_0+a_4x+. . . + a_{252}x^{63}$, $A_1(x)= a_1+a_5x+. . . + a_{253}x^{63}$, . . , $A_3(x)= a_2+a_6x+. . . + a_{255}x^{63}$, and resulting polynomial can be stored reduced by $x^{64}+1$ reducing the memory requirement to half with respect to most efficient software Toom-Cook implementation \cite{bermudo2020time}). Optimization of multiplication block (most power-hungry block) results in 1.37X power-efficient, 1.75X lower area and 4X less memory requirement w.r.t. other SoA PQC ASIC \cite{banerjee20192}. Our point multiplication is ~1.58X energy efficient than state-of-the art NTT based multiplier design, which proves strongest candidacy of Saber in NIST PQC competition. \\
It should be noted that previous implementations as of now are either for LWE based scheme\cite{song2018leia}, or post silicon is not verified \cite{zhu2021lwrpro}, \cite{cryptoeprint:2018:682}. Fig.3 explains the overview of the module that performs vector-vector multiplication in our accelerator. This module is micro-programmed to perform the vector-vector polynomial multiplication in Saber using a single polynomial signed multiplier. From an algorithmic point of view, the polynomial multiplier provides the first hardware implementation of the striding version of Toom-Cook multiplication, which is a compact variant of the well-known Toom-Cook multiplication. Detailed data path for evaluation, multiplication \& interpolation is presented in Fig.3. The evaluation data path is equivalent to other Toom-Cook versions \cite{mera2020compact}. In addition, we also use the lazy interpolation technique which helps in reducing latency, hence reducing energy consumption, and improving performance. 7 parallel point-multiplier with local memory (utilized by distributed SRAM in the chip micrograph) significantly improves the performance (7X latency improvement) at the expense of negligible accelerator area. This memory needs to allocate 2 polynomials of 64 coefficients X16 bits each (A \& C), 1 polynomial of 64 coefficients of 8 bits each (B) with dual port reading capabilities for faster approach. The point-multiplier can instantiate several multiply-and-accumulate (MAC) units to further parallelize the point-value multiplication. Using 4 multipliers gives similar performance to [4] while providing improvement in power, area, and energy. 2) For achieving the memory efficiency offered by striding Toom-Cook, the polynomial multiplication becomes a convolution where the expanded result is negatively wrapped. This nega-cyclic convolution can be achieved by performing the complement of two every time a coefficient that would lie out of the bounds shall be used. To do so, accounting for the fact that the length of the polynomials is 64, the sixth bit of the coefficient index can be used to perform or not the complement of two operation of every coefficient being read or written from or to memory, while only the lesser significant bits of the coefficient index are used to generate the memory address. 3) This highly area, energy and memory-efficient accelerator can be integrated with bigger SoCs/ resource constraint IoT devices readily using the high-speed serial interface. Interactive instruction set based architecture re-uses basic building blocks to reduce area overhead and makes it compatible with processors running in SoC. The 24-bit custom instruction set supports 4-bit opcodes (to support 11 operations) and two 10-bit address offsets to access 8KB data memory (dual port 1K address x 64-bit data bus). Binomial sampler is implemented using combinational logic as it does not incur high area.

\begin{figure*}[!t]
\centering
\includegraphics[width=1.4\columnwidth]{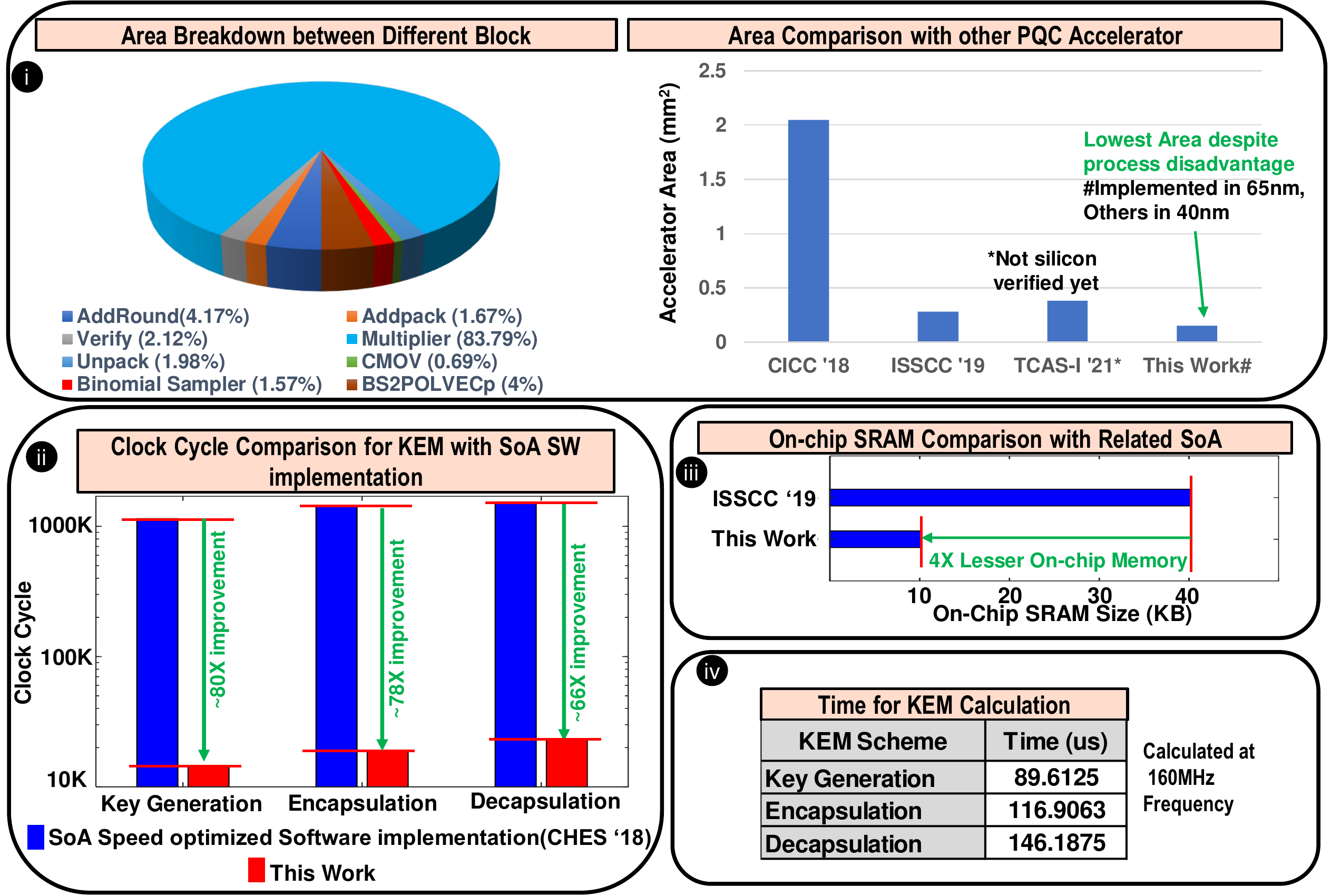}
\caption{i) Area breakdown \& comparison with State-of-the-art PQC implementation; ii) Clock cycle comparison with fastest SoA implementation on Cortex-M4; iii) On-chip memory comparison with SoA Learning With Error (LWE) PQC algorithms; iv) Time for KEM calculation.}
\label{voltage_glitch_waveform}
\vspace{-4mm}
\end{figure*}

\begin{figure}[!t]
\centering
\includegraphics[width=0.9\columnwidth]{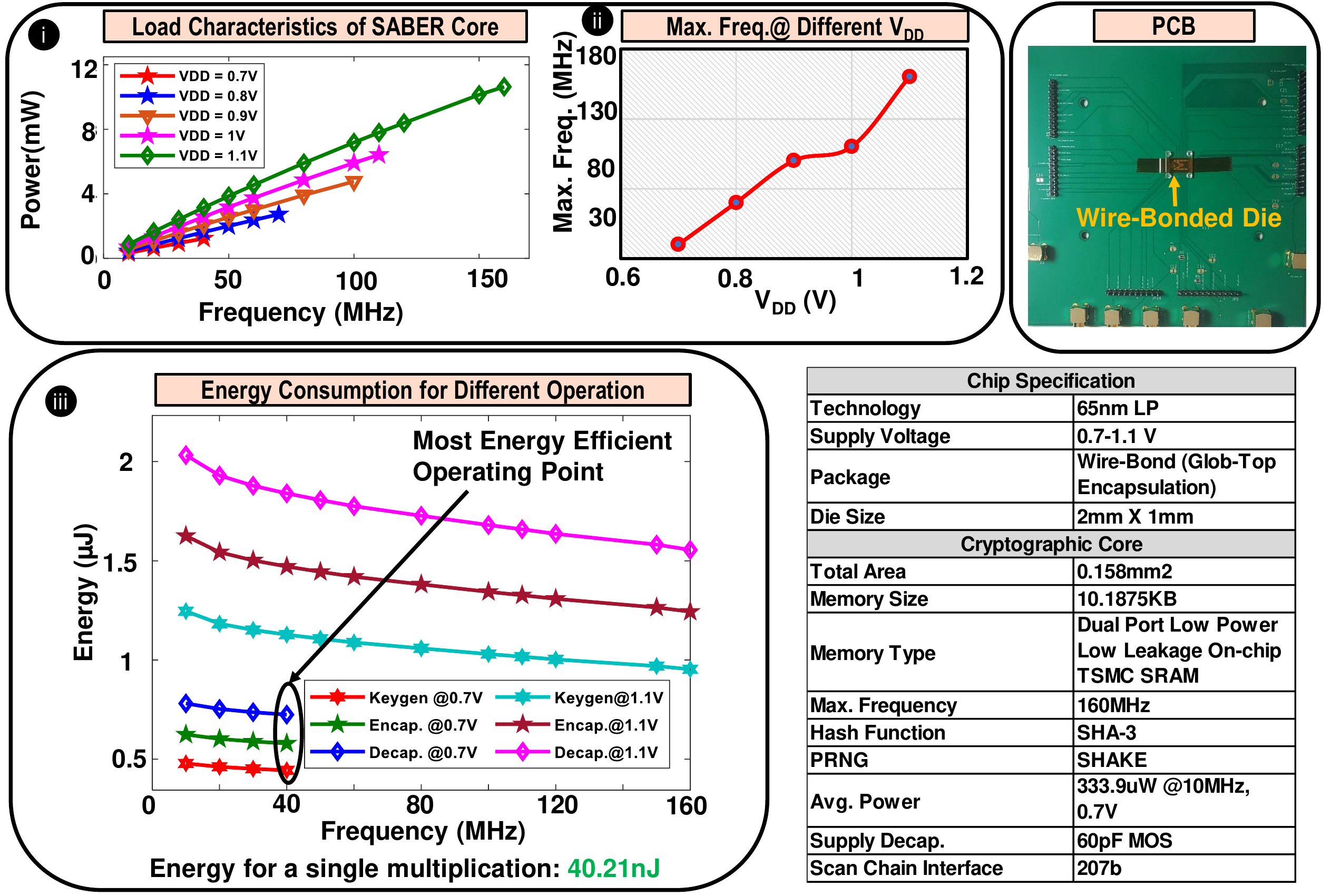}
\caption{i) Load Characteristics of Saber; ii) Maximum frequency; iii) Energy consumption; PCB \& Chip specification.}
\label{attack_detector}
\end{figure}

\vspace{-1mm}
\section{Measurement Results}
\vspace{-1mm}
Fig. 4 shows area breakdown and area comparison with SoA PQC accelerator. This accelerator consumes 1.75X lowest area with respect to other PQC accelerators (40nm) despite the disadvantages of process (65nm). Clock cycle requirement for Key Generation, Encapsulation and Decapsulation is 80X, 78X and 66X improvement with respect to SoA software. Striding memory access approach reduces the memory requirement to ~10KB, 4X lower than other SoA PQC core [4]. All the mechanism of KEM (keygen, encapsulation and decapsulation) are observed to be finished within 89, 117 \& 146 us respectively (~80X improvement over fastest software implementation, no hardware implementation till date). Power and energy consumption are presented in Fig. 5. Highly pipelined architecture results in maximum frequency of 160MHz at 1.1V VDD. Full core consumes only 334uW at 10MHz and 0.7V while being fully operational. It consumes the lowest energy at 40MHz and 0.7V VDD.  Key generation, encapsulation and decapsulation consume 444.1, 579.4 \& 724.5 uJ energy for entire operation in the above-mentioned operating point. A comparison table with other state-of-the-art implementations \cite{song2018leia}, \cite{zhu2021lwrpro}, \cite{cryptoeprint:2018:682} (LWE based) is presented in Fig. 6. This is the first silicon-verified implementation of LWR PQC core. ~1.58X lesser energy consumption of unit multiplication with respect to SoA NTT multiplication proves the credibility of Saber (LWR PQC) as a finalist of NIST PQC competition final round.  

\begin{figure}[!t]
\centering
\includegraphics[width=0.45\textwidth]{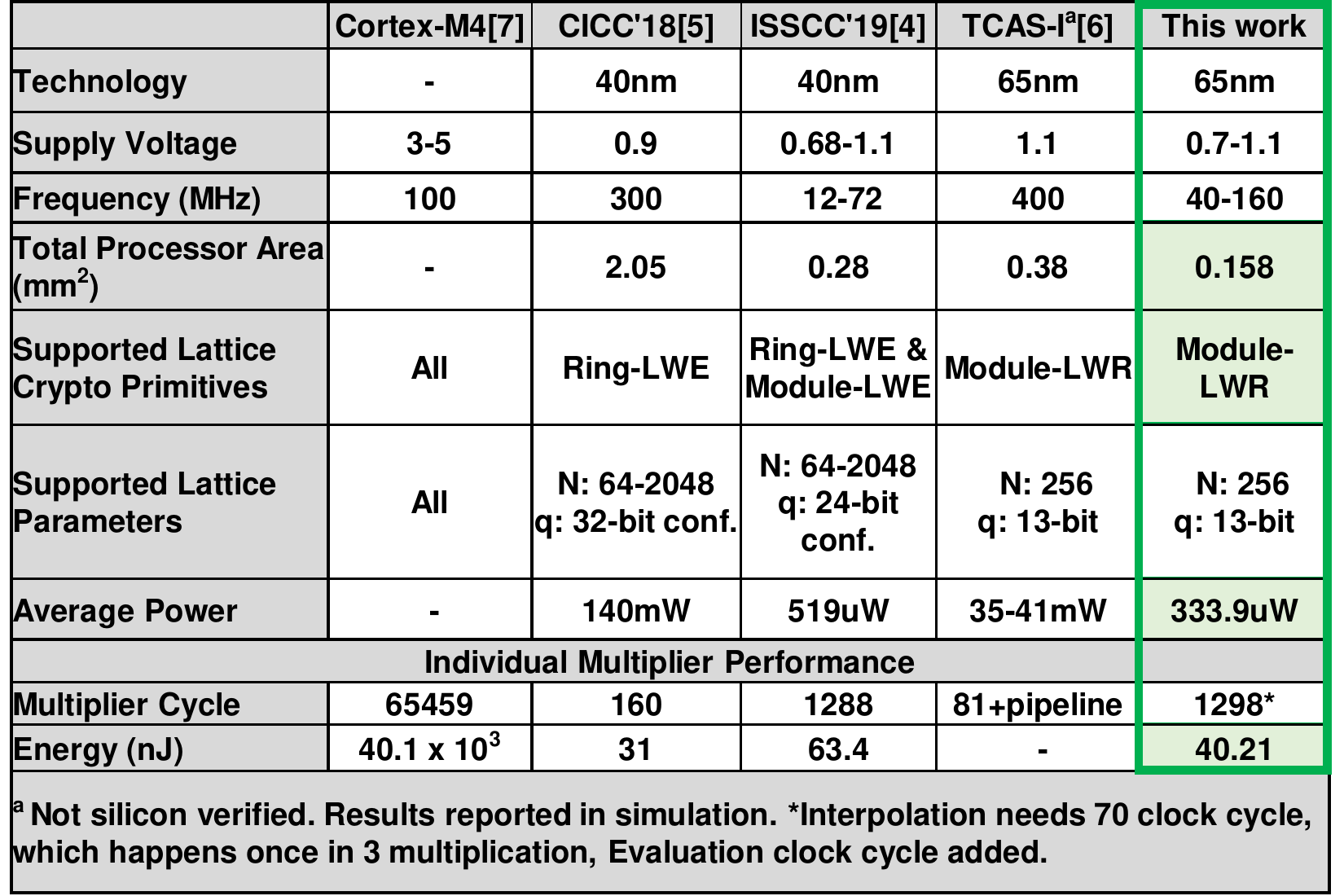}
\caption{Comparison of 1st Saber ASIC with Cortex-M4 software and other hardware post-quantum cryptography accelerators.}
\label{noise_resilience_time_domain}
\vspace{-3mm}
\end{figure}
\begin{figure}[!t]
\centering
\includegraphics[width=0.45\textwidth]{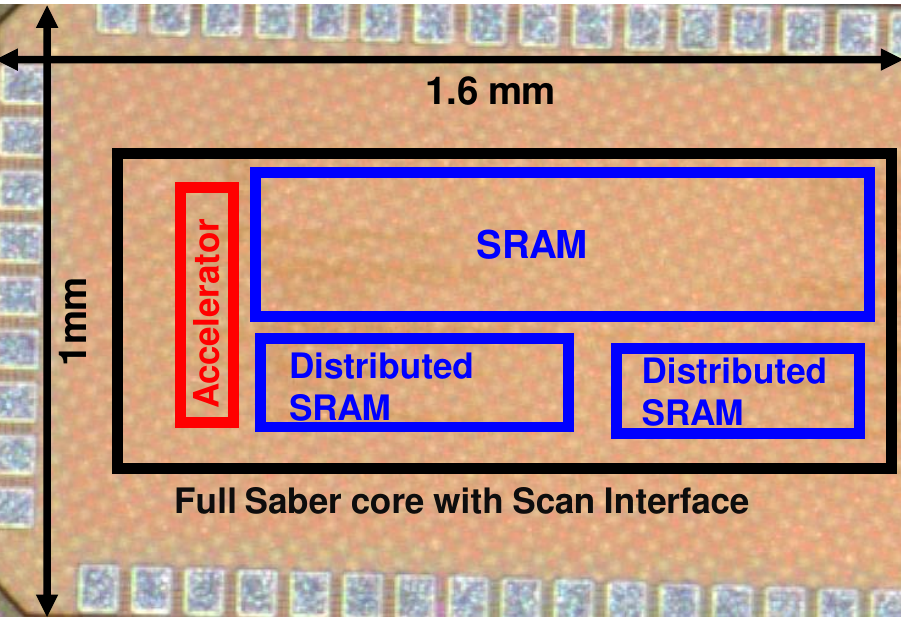}
\caption{Chip micrograph}
\label{noise_resilience}
\vspace{-5mm}
\end{figure}

\section*{Acknowledgement}
This work was partly supported by NSF (Grant CNS 17-19235), and Intel Corporation.

\bibliographystyle{unsrt}
\begin{small}
{
    \bibliography{main.bib}
}
\end{small}

\end{document}